\documentclass[]{spie}  

\usepackage{amsmath,amsfonts,amssymb}
\usepackage{graphicx}
\usepackage[colorlinks=true, allcolors=blue]{hyperref}
\usepackage{caption}
\usepackage{subcaption}
\usepackage{multirow}
\usepackage{siunitx}
\usepackage{xcolor}

\usepackage{xspace}

\def\spider{{\sc Spider}\xspace}

\def\SuperBIT{{SuperBIT }}

\def\taurus{Taurus}

\newcommand{\ukrts}{ $\mu\mathrm{K}_{\mathrm{\mbox{\tiny\sc cmb}}}\sqrt{\mathrm{s}}$ }

\def\gev{GeV~c$^{-2}$}

\def\gevpercm3{\gev~\percm3}
\def\gpercm3{g~\percm3}

\title{Instrument Overview of Taurus: A Balloon-borne CMB and Dust Polarization Experiment}

\author[a]{Jared L. May}
\author[b]{Alexandre E. Adler}
\author[c]{Jason E. Austermann}
\author[d]{Steven J. Benton}
\author[a]{Rick Bihary}
\author[g,c]{Malcolm Durkin}
\author[c]{Shannon M. Duff}
\author[e]{Jeffrey P. Filippini}
\author[d]{Aurelien A. Fraisse}
\author[f]{Thomas J.L.J. Gascard}
\author[e]{Sho M. Gibbs}
\author[d]{Suren Gourapura}
\author[f, b]{Jon E. Gudmundsson}
\author[h]{John W. Hartley}
\author[c]{Johannes Hubmayr}
\author[d]{William C. Jones}
\author[h]{Steven Li}
\author[a]{Johanna M. Nagy}
\author[c,g]{Kate Okun}
\author[a]{Ivan L. Padilla}
\author[h]{L. Javier Romualdez}
\author[d]{Simon Tartakovsky}
\author[c]{Michael R. Vissers}

\affil[a]{Department of Physics, Case Western Reserve University, 10900 Euclid Ave, Cleveland, OH 44106, USA}
\affil[b]{The Oskar Klein Centre, Department of Physics, Stockholm University, AlbaNova, SE-10691 Stockholm, Sweden}
\affil[c]{National Institute of Standards and Technology, 325 Broadway Mailcode 817.03, Boulder, CO 80305, USA}
\affil[d]{Department of Physics, Princeton University, Jadwin Hall, Princeton, NJ 08544, USA}
\affil[e]{Department of Physics, University of Illinois Urbana-Champaign, 1110 W Green St, Urbana, IL 61801, USA}
\affil[f]{Science Institute, University of Iceland, 107 Reykjavik, Iceland}
\affil[g]{Department of Physics, University of Colorado Boulder, Boulder, Colorado, USA}
\affil[h]{StarSpec Technologies Inc., Unit C-5, 1600 Industrial Avenue, Cambridge, ON N3H 4W5, Canada}

\authorinfo{Further author information: (Send correspondence to Jared May)\\Jared May: E-mail: jlm299@case.edu}

\begin{document} 

\maketitle

\begin{abstract}
\taurus{} is a balloon-borne cosmic microwave background (CMB) experiment optimized to map the $E$-mode polarization and Galactic foregrounds at the largest angular scales ($\ell <$ 30) and improve measurements of the optical depth to reionization ($\tau$). This will pave the way for improved measurements of the sum of neutrino masses in combination with high-resolution CMB data while also testing the $\Lambda$CDM model on large angular scales and providing high-frequency maps of polarized dust foregrounds to the CMB community. These measurements take advantage of the low-loading environment found in the stratosphere and are enabled by NASA’s super-pressure balloon platform, which provides access to 70\% of the sky with a launch from Wanaka, New Zealand.  Here we describe a general overview of \taurus{}, with an emphasis on the instrument design. \taurus{} will employ more than 10,000 100~mK transition edge sensor bolometers distributed across two low-frequency (150, 220~GHz) and one high-frequency (280, 350~GHz) dichroic receivers. The liquid helium cryostat housing the detectors and optics is supported by a lightweight gondola. The payload is designed to meet the challenges in mass, power, and thermal control posed by the super-pressure platform. The instrument and scan strategy are optimized for rigorous control of instrumental systematics, enabling high-fidelity linear polarization measurements on the largest angular scales. 
\end{abstract}


\section{INTRODUCTION}
\label{sec:intro}

\begin{figure}
    \centering
\includegraphics[width=9cm]
{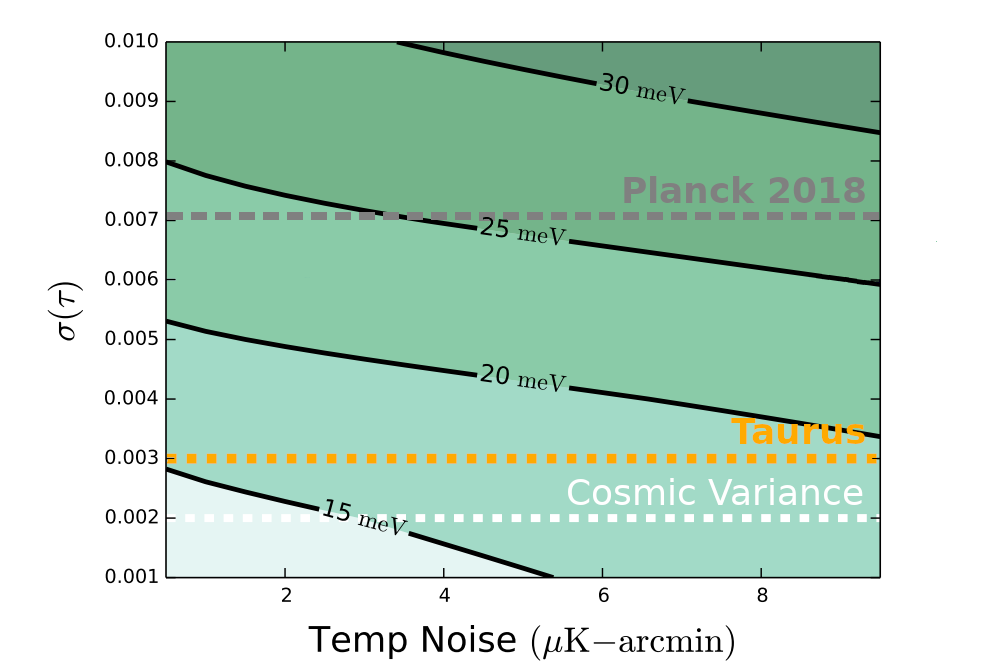}
\includegraphics[width=8cm]
{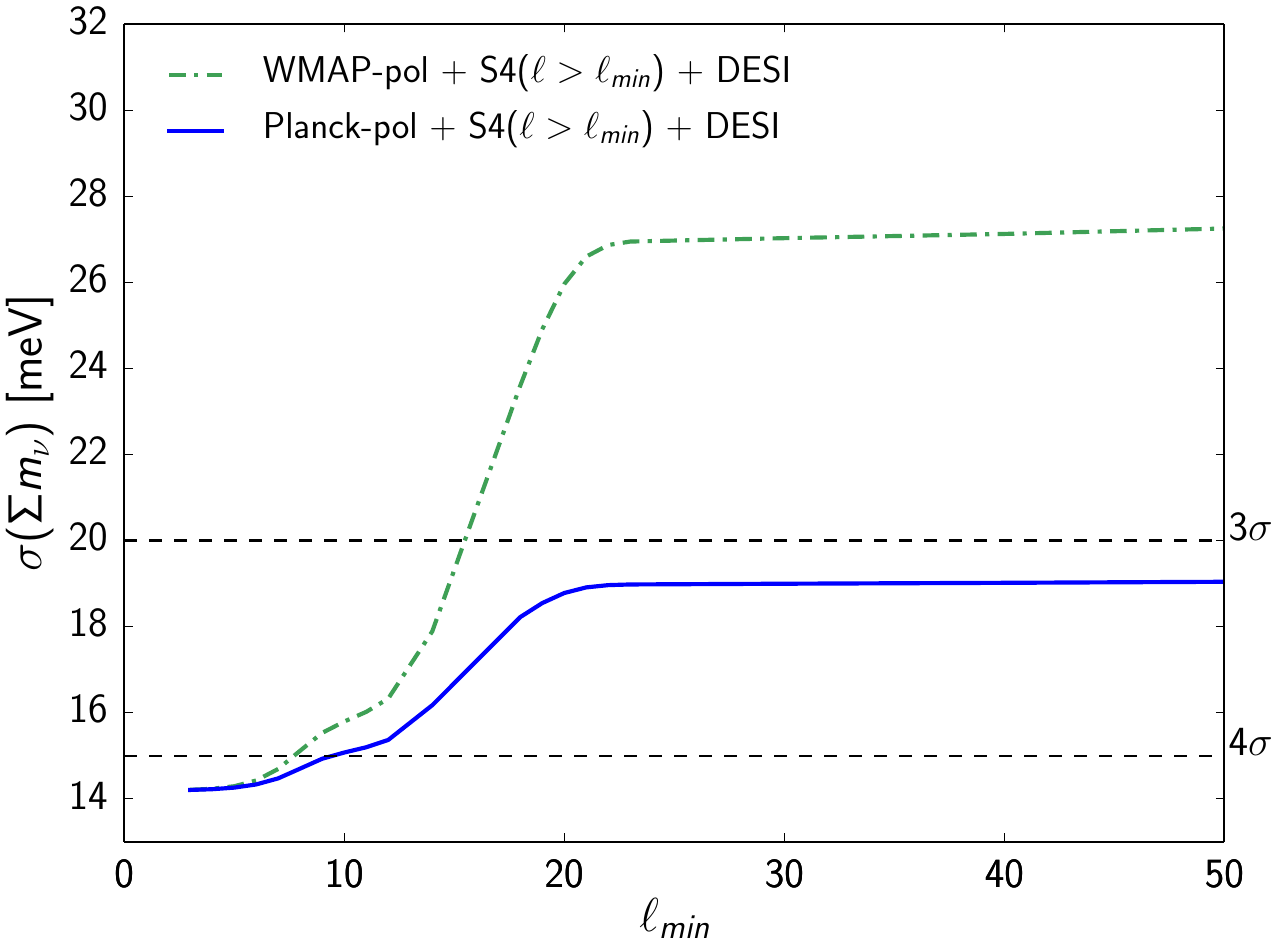}
    \caption{Left: Measurements of $\sum m_\nu$ (green contours) from high-resolution CMB maps with noise $<$ 10~$\mu K$-arcmin are nearly saturated by the degeneracy with current $\tau$ measurements \cite{abazajian2016cmbs4}. Right: However, improving $E$-mode polarization measurements at multipoles below $\ell \sim 20$ where the reionization signal dominates, can break this degeneracy and enable significant improvements on the $\sum m_\nu$ measurement.\cite{Allison_2015}}
    \label{fig:sigma_m_nu}
\end{figure}

The $\Lambda$CDM model of cosmology describes many features of the observable Universe with six parameters. Of these, the optical depth to reionization, $\tau$, is the least well constrained, with the fractional uncertainty almost an order of magnitude greater than the other five parameters\cite{2020}. The \textit{Planck} 2018 results report $\tau = 0.054 \pm 0.007$\cite{2020}, but more precise measurements of the cosmic microwave background (CMB) polarization on the largest angular scales have the potential to reduce this uncertainty.  As illustrated in Figure~\ref{fig:sigma_m_nu}, current and planned CMB lensing measurements of $\sum m_\nu$, the sum of neutrino masses, are strongly degenerate with $\tau$. \cite{act_neutrinos, spt-3g, Ade_2019, abazajian2016cmbs4}. Improved $\tau$ measurements are thus crucial for achieving tighter constraints on $\sum m_\nu$, potentially enabling the normal and inverted neutrino mass hierarchies to be distinguished.  Additionally, improved low-multipole ($\ell$) CMB power spectrum measurements would provide a valuable test of the $\Lambda CDM$ model, including the lack of correlations on large angular scales, the hemispherical power asymmetry, and the alignment
or excess amplitude of odd-parity modes in the temperature anisotropies \cite{2016Planck, o2020hemispherical}.

One major challenge in measuring CMB polarization on large angular scales is that the cosmological information is overshadowed by Galactic foregrounds.  At frequencies above 150~GHz, the polarized foregrounds are dominated by thermal emission from interstellar dust.\cite{Planck_dust} Although the dust amplitude varies by sky region, minimizing the uncertainty from sample variance in the low-$\ell$ CMB measurements requires observing as much of the sky as possible, necessitating robust component separation even in relatively dusty regions.  The deep high-frequency sky maps required for $\tau$ measurements would in turn provide a valuable resource to the CMB community, supplementing existing \textit{Planck} HFI maps\cite{Planck_dust}.

Here we present an overview of \taurus{}, a mid-latitude super-pressure balloon mission to map the polarization of the microwave sky with high fidelity at large angular scales. \taurus{} will measure the CMB and Galactic dust emission over $\sim$70\% of the sky in four frequency bands from $150 - 350$~GHz, where the advantage of the balloon platform relative to terrestrial sites is greatest.  These measurements are highly complementary to the lower frequency large angular scale measurements from the ground-based CLASS experiment \cite{CLASS} and can provide sky maps to the community before the planned LiteBIRD satellite \cite{litebird}. A combination of high sensitivity and systematics mitigation will enable \taurus{} to measure $\tau$ to within a factor of two of the cosmic variance limit from the low multipole $E$-mode polarization CMB power spectrum.

\section{Flight Plan}

\begin{figure}
    \centering
    \includegraphics[width=0.7\textwidth]{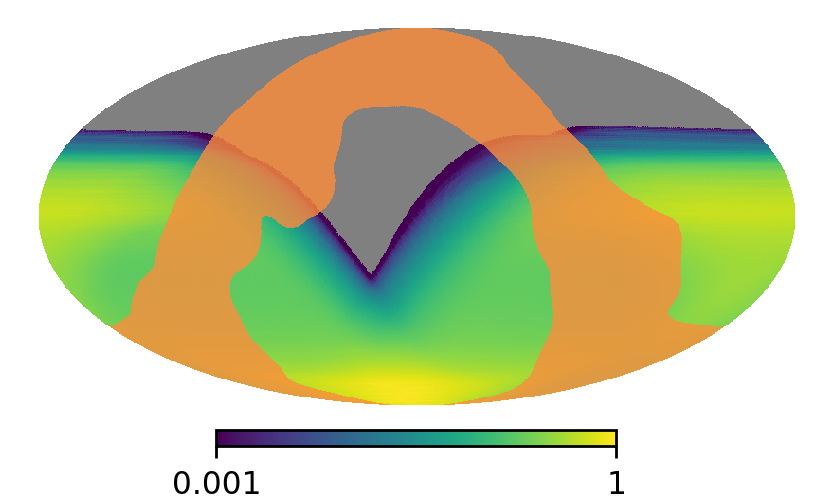}
    \caption{A simulated normalized \taurus{} hits map in equatorial coordinates assuming a late March launch and one month of nightly observations. The orange region shows a rough outline of the Galactic plane (from the Planck 2015 GAL070 mask), and the gray pixels fall outside \taurus{}' field of view.}
    \label{fig:hitsmap}
\end{figure}

\taurus{} is designed to observe from a Super-Pressure Balloon (SPB) flight launched from Wanaka, New Zealand at a latitude of \SI{44.7}{\degree} South. Spinning in azimuth at \SI{30}{\degree\per\second} at a fixed elevation of \SI{35}{\degree} will allow it to map $\sim$70\% of the sky.  Figure~\ref{fig:hitsmap} shows the simulated sky coverage after 1~month of observing time assuming a late March launch.  The scan rings cross a given point on the sky multiple times while both rising and setting, ensuring each map pixel is observed by multiple detectors at multiple orientations.  This cross-linking enables robust reconstruction of large-scale features, and alternating spin directions each day provides additional systematics mitigation.  Due to power limitations, \taurus{} plans to observe only at night, and with daylight hours spent charging the batteries, recycling the cryogenic refrigerators, and transmitting data.

\begin{figure}
    \centering
    \includegraphics[width=0.8\textwidth]{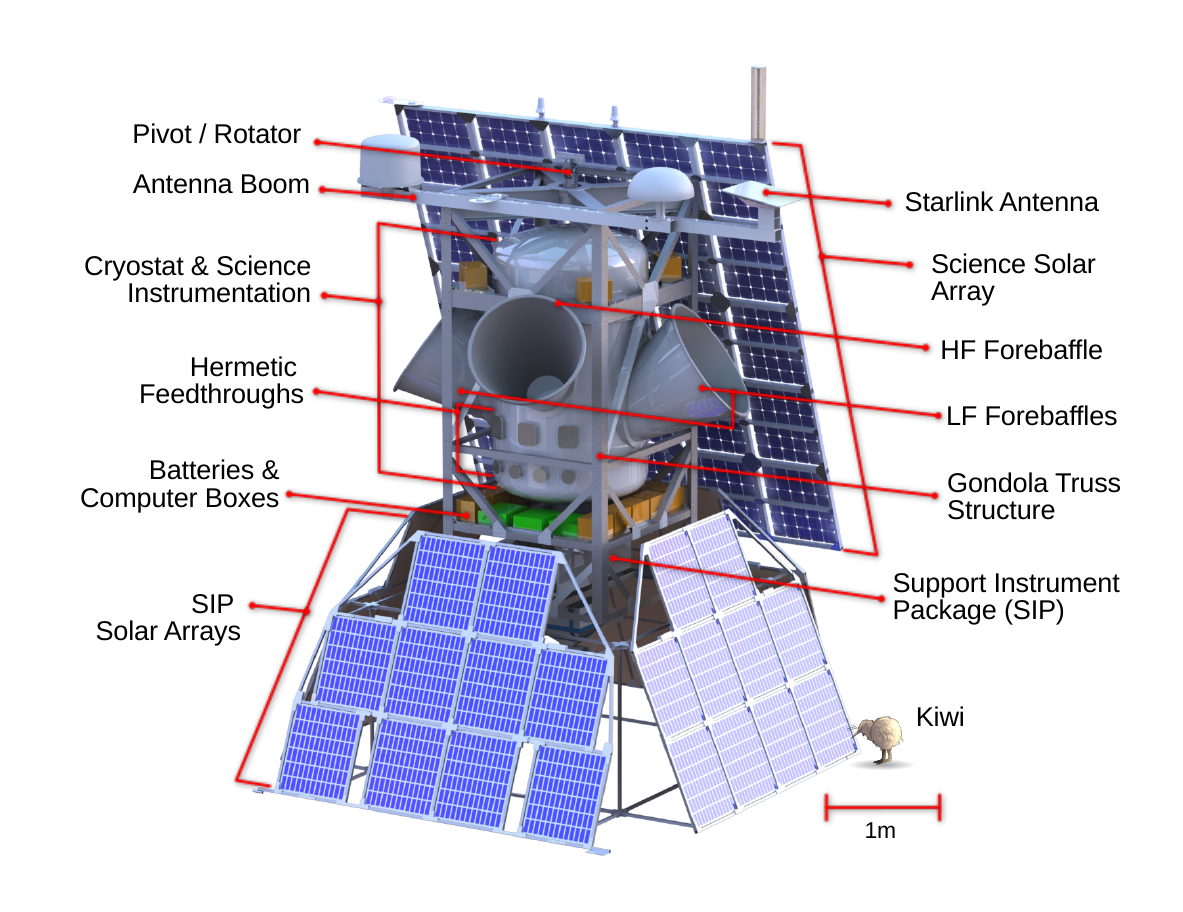}
    \caption{ A model of the \taurus{} payload. A rigid truss structure provides support for the cryostat, electronics, and SPB launch provider components, which include the Support Instrumentation Package (SIP), the SIP solar arrays, and communications antennas. The payload sun shielding and other thermal treatments are not drawn in order to more clearly show the main subsystems.}
    \label{fig:taurus_full_render}
\end{figure}

The \taurus{} payload consists of three refracting telescopes housed in a shared liquid helium cryostat supported by a rigid gondola. The total target mass of the science payload is $\leq 1500$~kg, and the overall design is summarized in Figure~\ref{fig:taurus_full_render} and Table~\ref{tab:instr_overview}. The main sub-systems are described in more detail in the sections that follow.

\begin{table}[h!]
\caption{A summary of the planned \taurus{} instrument design. Each spatial pixel consists of four science detectors (two polarization orientations at each of two frequency bands), and the instrument sensitivities assume 85\% end-to-end detector yield.}
\begin{center}
\begin{tabular}{|c|c|c|c|c|c|c|}
\hline
Band  &  & Beam & Number &  Absorbed & Detector & Instrument \\
Center  & Bandwidth & FWHM & of & Power & Sensitivity & Sensitivity \\
(GHz) & (GHz) & (arcmin) & Detectors & (pW) & (\ukrts)& (\ukrts) \\ \hline
150 & 40 & 30  & 3024 & 0.9 & 76 &  1.5 \\ 
220 & 55 & 22  & 3024 & 1.1 & 123 &  2.4 \\ 
280 & 70 & 26  & 2016 & 1.4 & 220 &  5.4 \\ 
350 & 85 & 22  & 2016 & 1.6 & 550 &  13.4 \\ 
\hline
\end{tabular}

\label{tab:instr_overview}
\end{center}
\end{table}

\section{Receivers and Cryogenics}

\subsection{Receiver Design}

\begin{figure}
    \centering
    \includegraphics[width=0.7\textwidth]{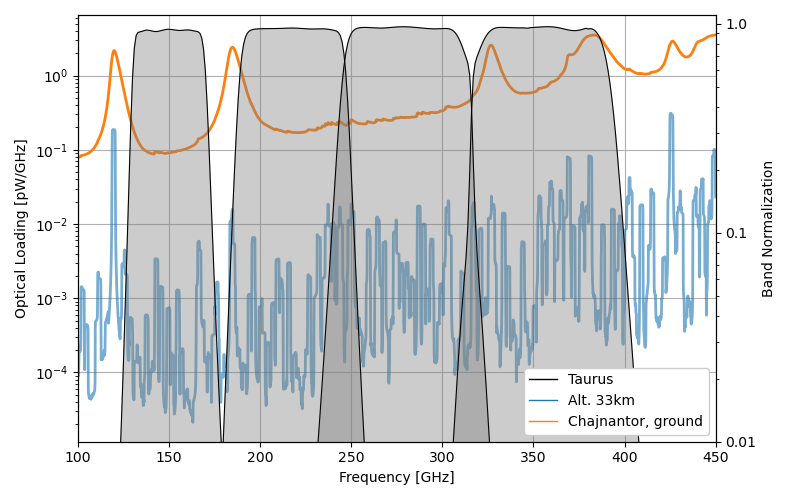}
        \caption{Estimates of \taurus's four observing bands are overplotted with the anticipated atmospheric loading from an altitude of 33~km.\cite{am}  For comparison, the loading from the Chilean Chajnantor plateau, a CMB observing site with access to similar sky coverage, is also shown.}
        \label{fig:LF_HF_BPs}
\end{figure}

\taurus{} will observe the sky with three dichroic receivers covering the frequency bands shown in Figure~\ref{fig:LF_HF_BPs}.  Two low-frequency (LF) receivers will measure bands centered on 150 and 220~GHz, and a single high-frequency (HF) receiver will measure 280 and 350 GHz bands. The \taurus{} receivers are being optimized to support the detector counts listed in Table \ref{tab:instr_overview} while maintaining high-quality beams and minimizing polarization systematics on the largest angular scales.  The azimuthally symmetric, on-axis design minimizes cross-polarization response and extensive internal and external baffling reduces beam sidelobe pick-up.  A more extensive discussion of \taurus{}'s beam systematics can be found in Adler et al.~2024\cite{Taurus_sys}. For additional systematics mitigation and to add redundancy to the scan strategy, the three \taurus{} receivers are azimuthally offset, with the two LF receivers separated by \SI{180}{\degree} and the HF halfway in between.  This introduces a phase shift to the sky signals they observe during the spin, rejecting thermal, microphonic, and magnetic systematics that would otherwise be in-phase for a co-pointed receiver design. 

Each receiver includes a stepped half-wave plate (HWP) polarization modulator mounted at 4~K.  Every day the HWPs will be stepped by integer multiples of \SI{22.5}{\degree}, covering a set of eight discrete angles in a repeating pattern over the course of the flight.  This complements the polarization modulation from sky rotation to provide complete $Q$ and $U$ Stokes parameter coverage on every detector \cite{hwp_nonidealities, vpol}.  The polarization-sensitive antennas on alternating spatial pixels are rotated by \SI{45}{\degree} with respect to each other, providing complete instantaneous polarization sensitivity for a given receiver.  The HWPs themselves will be anti-reflection-coated birefringent single-crystal sapphires. A multi-plate achromatic design will provide good performance over both bands of the dichroic receivers, as has been previously demonstrated by several other experiments \cite{blastpol_hwp, ebex_hwp, SO_HWP}.  The rotation mechanisms will be based on the successful \spider design, using stepper motor-driven worm gears operated at 4~K \cite{bryan_hwp, bryan_mechanism}.  A combination of absolute and relative encoders will allow post-flight angle reconstruction to $<\,$\SI{0.1}{\degree}.

\subsection{Cryogenic System}
The \taurus{} cryogenic system cools both the receiver optics and detectors and is described in more detail in Tartakovsky et al.~2024.\cite{Taurus_cryo}.  The cryostat itself is a lightweight aluminum vacuum vessel that draws heavily from the heritage of the \spider design \cite{Gudmundsson_2010,Gudmundsson_2015}. A 660~L tank of liquid $^4$He provides the primary source of cooling power.  Cooling below 2~K is provided by a 5~L tank of superfluid $^4$He, which is continuously filled through a capillary system.  While at float, the pressure of this tank is maintained by the ambient stratospheric pressure.  In addition to cooling the internal receiver baffles, the superfluid tank also provides cooling power to the condensation point of the closed-cycle $^3$He refrigerators, which achieve base temperatures $<300$~mK.  These in turn provide cooling power to the condensation point of Chase Research Cryogenics miniature dilution refrigerators\cite{Brien_2018}, which cool the detectors to $\leq100$~mK. To achieve the long hold time desired for an SPB flight, the inner stages of the cryostat are surrounded by two vapor-cooled shields, which use the available enthalpy of the $^4$He boil-off from the main tank.  Infrared (IR) filters at each temperature stage reduce the out-of-band loading, using a combination of lossy plastic and metal-mesh low-pass filters\cite{Cardiff_filters}.  

\section{Detectors and Readout}
The \taurus{} detectors are large-format arrays of dichroic horn-coupled transition edge sensor (TES) bolometers \cite{Duff_2016}, with the detector counts listed in Table~\ref{tab:instr_overview}. These arrays are based on an architecture developed by NIST that has been successfully deployed by a number of experiments \cite{kusaka2018,austermann2012,thornton2016,dicker2014,hubmayr2016,henderson2016,so_forecast2019}. Each focal plane is built from sub-units consisting of an array of feedhorns mated to a silicon detector assembly.  Each horn delivers radiant power to a planar orthomode transducer (OMT), which couples polarized light onto planar superconducting circuits that further distribute signals to diplexers and TES bolometers for power sensing within two bands and two linear polarizations. The TES bolometers consist of thin films of Al-Mn ($T_c\approx 160$~mK) thermally isolated by narrow legs of silicon nitride.  

\begin{figure}
\captionsetup[subfigure]{labelformat=empty}
    \centering
    \begin{subfigure}[b]{0.4\textwidth}
        \centering
        \raisebox{+0.4\height}{\includegraphics[width=\textwidth]{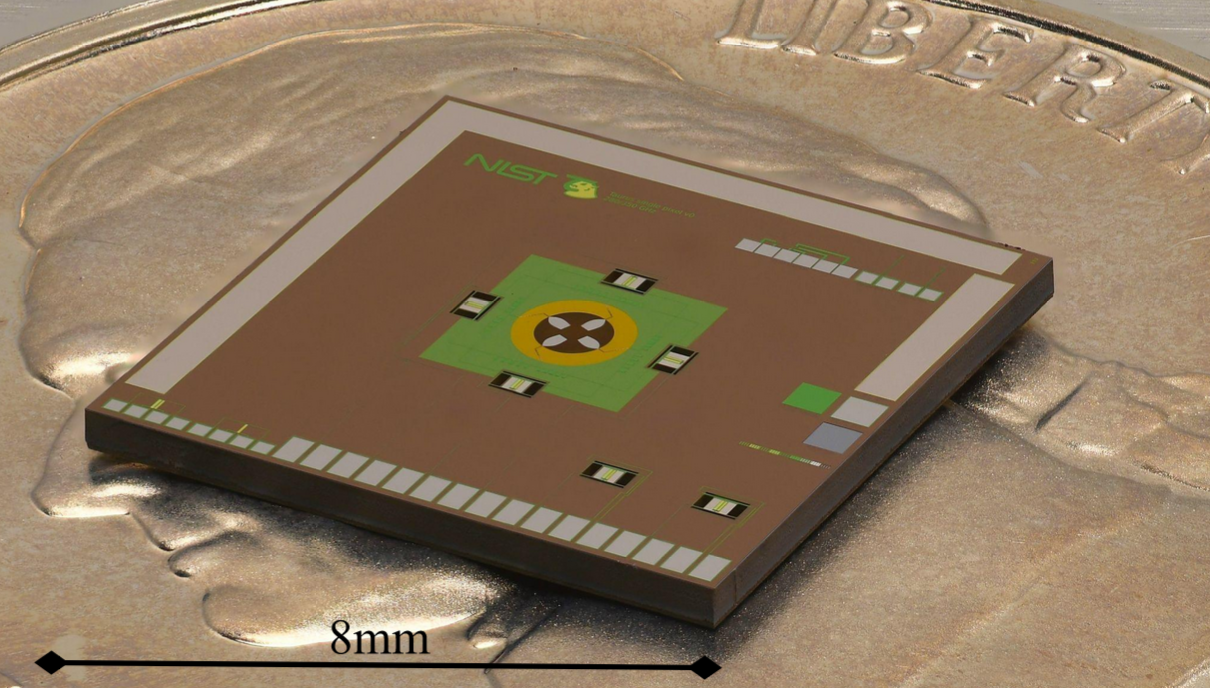}}
        \caption{}
        \label{fig:LF_dets}
    \end{subfigure}
    \begin{subfigure}[b]{0.5\textwidth}
        \centering
        \includegraphics[width=\textwidth]{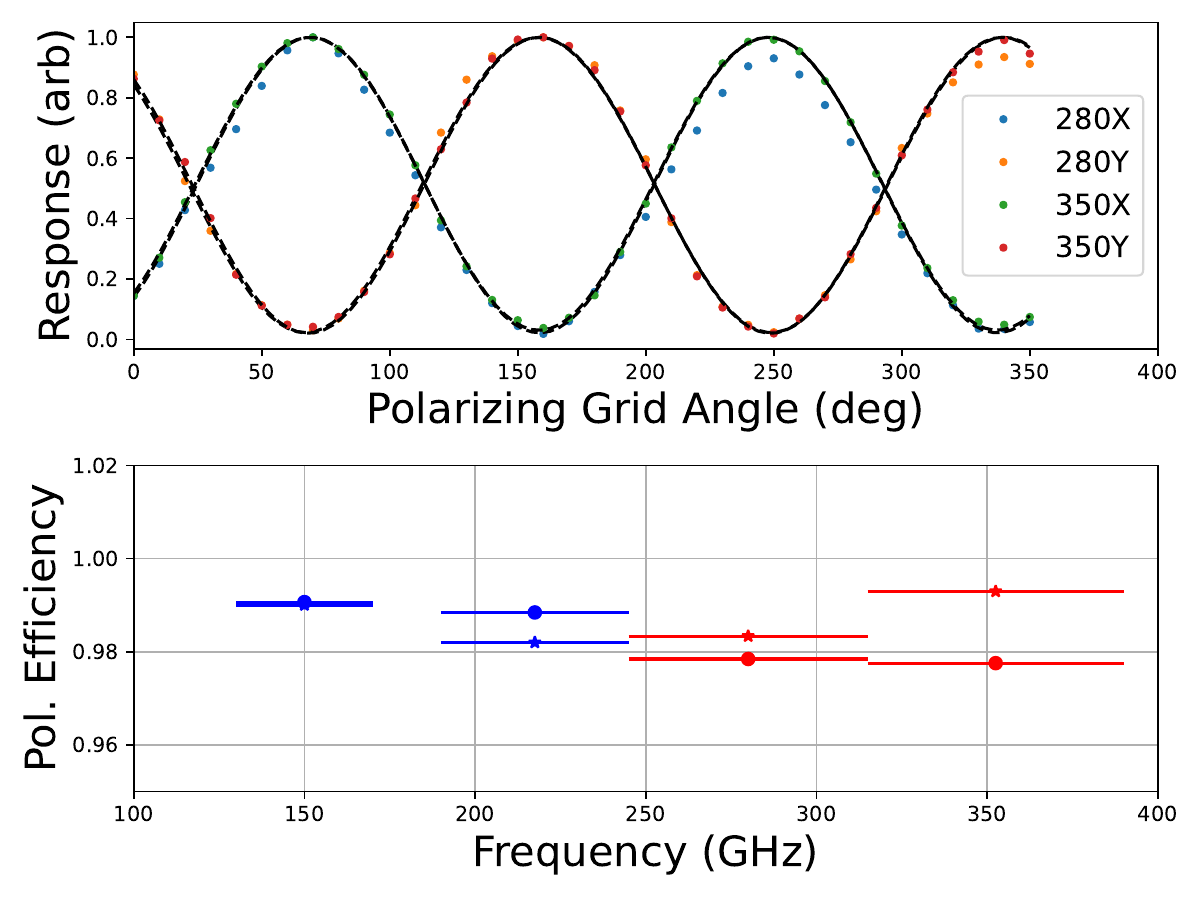}
        \caption{}
        \label{fig:HF_dets}
    \end{subfigure}
        \vspace{-12pt}
        \caption{Left: Photograph of a 280/350~GHz dual-band, dual-polarization device with specifications optimized for \taurus{}.  
        Both the 150/220~GHz and 280/350~GHz pixel-types have been designed, fabricated at NIST, and characterized.  
        Right: Verified polarization sensitivity.   
        Top: Response of all four bolometers within the 280/350~GHz device to a polarized source.  The dashed black lines are fits to a sine wave.  Bottom: The measured polarization efficiency is near or exceeds 98\% in both bands and polarizations for both \taurus{} pixel-types (blue: 150/220~GHz, red: 280/350~GHz).  Circles (stars) denote the X (Y) detector polarization response.  The horizontal bars for each data point show the designed passband.}
        \label{fig:LF_HF_dets}
\end{figure}

\taurus{}-optimized dichroic devices at 150/220~GHz and 280/350~GHz have been fabricated and characterized.  The measured saturation powers for all four bands are within 10\% of the design targets, and as such the dynamic range and sensitivity requirements should be satisfied by these detector realizations.  
The 350~GHz band is the highest NIST OMT-coupled frequency band characterized to date. 
The top right sub-figure of Figure~\ref{fig:LF_HF_dets} shows the response of the 280/350~GHz device to an on-axis polarization source.  The bolometer response is well-fit to a simple sinusoid (black dashed lines) for all four signal channels, and each X-Y polarization pair is phase-shifted by \SI{90}{\degree} as expected.  The polarization efficiency (left, bottom of Figure~\ref{fig:LF_HF_dets}), defined as $1-\min{(R)}/\max{(R)}$ from the sinusoidal fit to the response data $R$, is near or exceeds 98\% in all cases.  
No unexpected trends with frequency or polarization-pair are observed.

\taurus\ is designed to take advantage of the very low-loading environment provided by the SPB platform. The corresponding low saturation power of
the Al-Mn TESs means that they do not function under room-temperature optical loading conditions; for pre-launch characterization, a secondary Al TES with higher saturation power (due to higher $T_c \approx 500$mK) is co-located and connected in series with the Al-Mn TES. These ``calibration'' TESs were used for the polarization measurements presented in Figure~\ref{fig:LF_HF_dets}.  Among the four TESs (two polarizations and two passbands) within each spatial pixel, their shared feed structure,  shared readout chain, and close proximity on the wafer efficiently suppress systematic differences that could result in false polarization signals.
 
To limit the readout wire count between the sub-K system and ambient temperature, \taurus{} employs a two-stage time-division SQUID multiplexer (TDM) readout system~\cite{Doriese16}.  
Readout development for \taurus{} closely follows that for the CMB-S4 project~\cite{DRM_SPIE,Goldfinger2024}, which supports similar per-wafer sensor counts using modern TDM developments~\cite{Durkin23}.
Variations in the current through each TES are detected by a dedicated first-stage SQUID ammeter. Each group of 64 first-stage SQUID channels (63 bolometers plus one ``dark'' channel to monitor amplifier drifts) is read out by a single series SQUID array and ambient-temperature amplifier chain. An array of Josephson junction-based switches is used to link a single channel at a time to this amplifier chain, such that each detector in the group is read out in rapid sequence through their common amplifier. 
Following packaging architectures developed for \spider and BICEP Array\cite{spider_instr10,Hui_BICEParray}, the SQUIDs are 
housed behind the detector tiles within layers of passive magnetic shielding. The individual SQUIDs and SQUID input coils are also wound as first and second-order gradiometers respectively, further limiting any impact from external magnetic fields.
Warm electronics to read out and control the TESs and SQUID multiplexer system will follow design work for CMB-S4, which is itself derived from the proven UBC Multi-Channel Electronics (MCE)\cite{Battistelli:2008mce}.

\section{Gondola and Pointing}

The cryostat and electronics are supported by the lightweight gondola frame, shown in Figure~\ref{fig:taurus_full_render}, and designed and constructed by StarSpec Technologies.\footnote{https://www.starspectechnologies.com/} 
Leveraging strong heritage in design, fabrication, certification, and flight qualification of balloon-borne gondolas, the Taurus gondola is designed to meet all Columbia Scientific Balloon Facility (CSBF) and NASA launch provider requirements for SPB, including those related to launch, chute shock, and landing. To simplify in-field operations, the \taurus{} gondola satisfies the constraints of SPB ground infrastructure while allowing for modular disassembly for shipping by means of standard truss elements. 
Azimuthal control, to enable the \SI{30}{\degree\per\second} scans at a fixed elevation of \SI{35}{\degree}, will be achieved with a linear torque motor in a pivot to dump angular momentum to the balloon over
longer time scales.  Similar technology has been used by BOOMERanG, BLAST, \spider, and SuperBIT \cite{shariff_pointing, romualdez2020robust}. No control is necessary in pitch or roll, keeping the \taurus{} gondola design simple and lightweight. Payload attitude will be determined to better than 1~arcmin post-flight using a combination of rate gyroscopes, a three-axis magnetometer, and redundant star cameras, all of which have strong flight heritage from previous experiments and have been successfully used to provide better pointing accuracy than is needed by \taurus{} \cite{gandilo_pointing, SuperBIT_pointing}.  During the daytime, pinhole sun sensors will also be used to keep the solar panels at their optimal charging orientation.  

Observing all night, \taurus{} must use as little power as possible so that its batteries can last until the next morning.  A battery bank of $\sim$170~kg lithium deep-cycle 48~V batteries is sufficient for \taurus{} to last for 16 hours.  Maintaining daytime operations and recharging batteries in under \SI{8}{\hour} requires a little over \SI{2}{\kilo\watt} from the solar panels.  All detector readout is disabled during the day to conserve power, and the gondola enters a simpler pointing mode keeping solar panels oriented relative to the sun.  The only extra power expenditure during the day, aside from charging batteries, is recycling the sub-K coolers, which only takes a fraction of the day and contributes $\ll$ \SI{1}{\watt} averaged over the whole day.

Surviving day-night cycles presents a significant thermal challenge for SPB payloads.  Sun shielding and other thermal treatments are designed to protect the cryostat, optics, and sensitive electronics from experiencing temperatures outside of their verified operating range.  Power is also budgeted for heating during the night, particularly to prevent the lithium batteries from cooling and losing capacity. 

During the daytime, \taurus{} will transmit data to the ground from the previous night. While work is underway to significantly increase the downlink bandwidth available on mid-latitude flights, at the current bandwidth of around 5~Mbps, \taurus{} would require lossy compression of its downlinked data. To mitigate the significant risk in the recovery of the payload and raw data, small data vessels could also be dropped while over land, as demonstrated by \SuperBIT \cite{data_drop, data_drop2}. Such vessels could easily carry the full uncompressed \taurus{} data set, and recovery within \SI{600}{\meter} of the predicted landing site was achieved by SuperBIT.

\section{Conclusion}

\taurus{} is a balloon-borne CMB polarimeter that will observe $\sim$ 70\% of the microwave sky from a mid-latitude super-pressure flight.  It is optimized to provide robust maps on the largest angular scales in four frequency bands centered at 150, 220, 280, and 350~GHz. This will improve measurements of $\tau$ from the CMB $E$-mode polarization, testing the $\Lambda$CDM model and breaking a parameter degeneracy in small-scale CMB measurements of the sum of neutrino masses. \taurus{} will also provide deep maps of the polarized Galactic foreground emission in a region of the sky being measured at lower frequencies by current and planned ground-based experiments\cite{abazajian2016cmbs4, Ade_2019}. As the first mm-wave instrument operating from the super-pressure balloon platform, \taurus{} will also serve as a technology demonstrator and provide valuable heritage for future experiments.

\acknowledgments     
 
\taurus{} is supported in the USA by NASA award number 80NSSC21K1957. JEG and TJLJG acknowledge support from The Icelandic Research Fund (Grant number: 2410656-051) and the European Union (ERC, CMBeam, 101040169).

\bibliography{report} 
\bibliographystyle{spiebib} 

\end{document}